\documentclass[aps,notitlepage,longbibliography]{revtex4-1}
\usepackage{graphicx}
\usepackage{amsmath,amsfonts,amssymb}
\usepackage{mathtools}
\usepackage{tikz}
\usepackage{placeins}
\usepackage{hyperref}
\usepackage[utf8]{inputenc}
\usepackage[T1]{fontenc}

\newcommand*{\e}{\mathrm{e}}
\newcommand*{\I}{\mathrm{i}}
\newcommand*{\nn}{\nonumber \\}
\newcommand*{\sst}[1]{{\substack{#1}}}
\newcommand*{\prodz}[2]{\prod_{k=1}^{#2} \sigma_{#1}^z(k)}
\newcommand*{\brprodz}[2]{\left[ \prod_{k=1}^{#2} \sigma_{#1}^z(k) \right]}
\newcommand*{\prodx}[2]{\left[ \prod_{k=1}^{#2} \sigma_{#1}^x(k) \right]}

\newcommand*{\vektor}[3]{\ensuremath{\left( \begin{smallmatrix} #1 \\ #2 \\ #3 \end{smallmatrix} \right)}}
\newcommand*{\up}{\;\!\uparrow\;\!}
\newcommand*{\dn}{\;\!\downarrow\;\!}
\newcommand*{\uket}{\left| \up \up \up \right\rangle}
\newcommand*{\dket}{\left| \dn \dn \dn \right\rangle}
\newcommand*{\ket}[1]{\left|#1\right\rangle}
\newcommand*{\ie}{\mbox{i.\,e.}}
\newcommand*{\eg}{\mbox{e.\,g.}}
\newcommand*{\hc}{\text{h.\,c.}}
\newcommand*{\lft}{\!\left}
\newcommand*{\figref}[1]{Fig.~\ref{#1}}

\begin{document}

\title{Jordan--Wigner transformations for tree structures}

\author{Stefan \surname{Backens}}
\affiliation{Institut f\"ur Theorie der Kondensierten Materie, Karlsruhe Institute of Technology,
D-76131 Karlsruhe, Germany}
\author{Alexander \surname{Shnirman}}
\affiliation{Institut f\"ur Theorie der Kondensierten Materie, Karlsruhe Institute of Technology,
D-76131 Karlsruhe, Germany}
\affiliation{Institute of Nanotechnology, Karlsruhe Institute of Technology, D-76344
Eggenstein-Leopoldshafen, Germany}
\author{Yuriy \surname{Makhlin}}
\affiliation{Condensed-matter physics Laboratory, National Research University Higher
School of Economics, 101000 Moscow, Russia} 
\affiliation{Landau Institute for Theoretical Physics, acad.~Semyonov av. 1a, 142432,
Chernogolovka, Russia}


\begin{abstract}
The celebrated Jordan--Wigner transformation provides an efficient mapping between spin chains and 
fermionic systems in one dimension. Here we extend this spin--fermion mapping to arbitrary
\emph{tree} structures, which enables mapping between fermionic and spin systems with
nearest-neighbor coupling. The mapping is achieved with the help of additional spins at the junctions 
between one-dimensional chains. This property allows for straightforward simulation of Majorana
braiding in spin or qubit systems.
\end{abstract}

\flushbottom
\maketitle

\thispagestyle{empty}

\section{Introduction}

The well-known Jordan--Wigner transformation relates spin-$\tfrac{1}{2}$ operators to fermionic
creation and annihilation operators. Thereby, it allows for mapping between spin and fermionic
systems. It was originally used by Jordan and Wigner to define fermionic operators in the
second quantization~\cite{JW}.
The Jordan--Wigner transformation introduces non-local ``string operators'' to
transform commuting operators of different spins into anti-commuting fermionic operators, and, in
general, does not preserve locality. Nevertheless, it maps even-parity local fermionic
Hamiltonians to local spin Hamiltonians; moreover, certain spin Hamiltonians in 1D are mapped to
free fermionic Hamiltonians, which are readily solvable~\cite{LiebSchultzMattis}.

Generalizations to higher dimensions were
discussed in recent decades~\cite{Fradkin,YWang,HuertaZanelli,SWang,VerstraeteCirac,OrtizBatistaAdvPhys}. 
They map (even-parity) fermionic Hamiltonians to
spin Hamiltonians, but even local quadratic fermionic terms are mapped onto operators which involve
many, in principle infinitely many, spin operators (though in some
cases~\cite{Fradkin,HuertaZanelli} the weight of the involved spins may slowly decay with distance).
One may also consider introducing
ancillary degrees of freedom. For instance, Verstrate and Cirac~\cite{VerstraeteCirac} suggested
doubling the number of degrees of freedom for a 2D lattice to achieve local, but not necessarily
simple Hamiltonians in the spin language.

Thus, free fermionic Hamiltonians are often mapped to complicated operators in the spin
language.
In Ref.~\cite{qsk}, a modified Jordan--Wigner transformation was proposed, such that a three-leg star
graph of free fermions (with nearest-neighbor hopping) could be mapped to a three-leg star graph of
spin chains (with nearest-neighbor couplings). The mapping required the introduction 
of an extra spin-$\tfrac{1}{2}$ in the vertex of the spin graph, coupled to the three spin chains
locally via a specific 3-spin coupling. Furthermore, in Ref.~\cite{qsk} an alternative
scenario was described, in which a three-leg spin graph with exclusively 2-spin interactions was
mapped to a Kondo-like system of fermionic chains coupled by one spin (cf. application in Refs.~\cite{Tsvelik,Pino}). 
Here, we demonstrate that these transformations can be generalised to binary-tree
structures of 1D chains, \ie\ connected, acyclic graphs with no more than three edges at each
vertex. Furthermore, we argue that this result can be directly generalized to generic, non-binary
trees.

This kind of transformation is of special interest in particular since it can be used to
simulate the physics and, notably, non-abelian statistics and braiding of fermionic
Majorana modes~\cite{ReadGreen,IvanovBraiding} in a
(topologically non-protected) spin system. 
For the case of a $T$-junction geometry with a single topological segment in the chain providing two
Majorana modes, this implementation was explicated in Ref.~\cite{emb}. Here we describe
braiding operations between Majorana modes belonging to different topological segments
in a system where the number of segments is arbitrary. 
A binary-tree structure may be viewed as consisting of many $T$-junctions; such structures may be
useful for implementation (physical simulation) of the Majorana braiding operation~\cite{1dw,Aguado}
with applications in
topological quantum computing.
We also argue that such spin systems mimic fermionic quantum
computers~\cite{BravyiKitaev00}, which can be efficient, \eg,
in quantum-chemistry simulations. Namely, braiding or other logic gates between remote qubits
naturally include Jordan-Wigner string operators, 
making these qubits fermionic. Encoding the population of molecular orbitals in 
such qubits (see e.g., Ref.~\cite{AspuruGuzik}) thus brings a considerable advantage for the
computing algorithms.

An explicit description of a Majorana braiding operation between two topological segments,
implemented in the corresponding spin system, is given in the Appendix.

\section{Geometry and notations}

\begin{figure}[hbt]
\newcommand*{\x}{1.4}
\newcommand*{\y}{0.7}
\newcommand*{\z}{0.9}
\newcommand*{\R}{0.12}
        \begin{tikzpicture}[>=latex]
        \path (0, 0) coordinate (c1)
                arc(315:75:\R) coordinate (c2)
                arc(75:195:\R) coordinate (c3);
        \draw (c1) -- (c2) -- (c3) -- (c1);
        \draw[->] (c2) -- ++(0, \x) node[above] {2};
        \draw[->] (c3) -- ++(-\x, 0) node[left] {3};
        \draw[->] (c1) node[above left] {$S_0^\beta$} -- ++(\y, -\y)
                node[above right] {1} -- ++(\y, -\y) coordinate (end1) node[below right] {$S_1^\beta$};
        \path (end1) arc(135:255:\R) coordinate (c11)
                arc(255:15:\R) coordinate (c12);
        \draw (end1) -- (c11) -- (c12) -- (end1);
        \draw[->] (c11) -- ++(0, -\x) node[below] {11};
        \draw[->] (c12) -- ++(\z, 0) node[above] {12} -- ++(\z, 0) coordinate (end12) node[right=1pt] {$\:S_{12}^\beta$};
        \path (end12) arc(180:300:\R) coordinate (c121)
                arc(300:60:\R) coordinate (c122);
        \draw (end12) -- (c121) -- (c122) -- (end12);
        \draw[->] (c121) -- ++(\z, -\z) node[right] {121};
        \draw[->] (c122) -- ++(\z, \z) node[right] {122};
        \end{tikzpicture}
\caption{Binary-tree structure consisting of fermionic or spin chains. 
In each chain, spins or fermions are numbered in the direction of the arrow. 
The descendent chains in the tree are identified by sequential numbers that are appended
to the label of their parent chain. 
The root is the only vertex with no incoming edge and may have three outgoing edges. 
The notations for the vertices and the chains are explained in the text and evident from the
diagram.
}
\label{struc}
\end{figure}
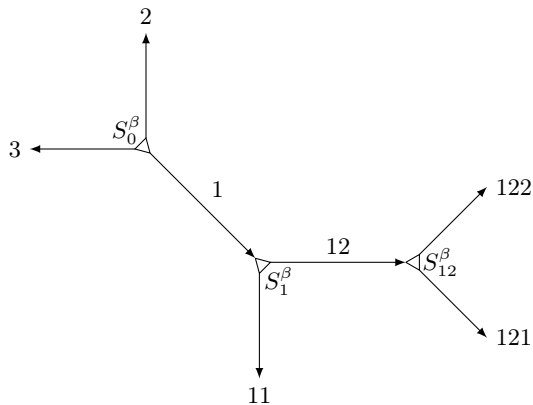

We consider spins on a tree-like lattice of the type depicted in \figref{struc}. Each edge of
the tree is a one-dimensional spin chain. The chains are connected at the vertices, and the
whole structure indicates the notion of locality (in fact, we focus on nearest-neighbor couplings).
In binary trees, they are connected in triples and, in general, interactions between boundary spins from all
three chains are allowed, so-called $\Delta$-junctions~\cite{Tsvelik}, indicating all
three pairwise couplings. In the particular case when one of the three couplings in the junction
vanishes, we obtain a $T$-junction, where all three chains have a common boundary spin.
We also consider fermions on the same tree and discuss methods to convert between spin and
fermionic systems.

A priori, the tree structures do not have a distinctive root and the edges do not
have orientation. 
For the purposes of the transformation, however, we choose an arbitrary vertex as a root
and assign to each edge (\ie\ chain) an orientation away from the root.
Based on this hierarchy, we introduce a notation for our further discussion by assigning a name to
each vertex and chain in the tree: The root is denoted ``0'' and the three outgoing chains acquire
numbers 1, 2, and 3. Then, step by step, each other vertex acquires a name $\alpha$, identical with
the incoming chain, while the two outgoing chains are assigned a longer name, $\alpha\beta$, with
$\beta=1$ or $2$, see \figref{struc}.

According to the orientation, the spins or fermions in each chain~$\alpha$ are numbered from 1 to
its length $L_\alpha$; they are represented by the Pauli matrices~$\sigma_\alpha^{x,y,z}(j)$
and the fermionic creation/annihilation operators~$c_\alpha^\dag(j) / c_\alpha(j)$, respectively.
To construct a fermion--spin transformation, we shall need ancillary spin operators, one per chain, which
we assign to the vertex at the beginning of the chain. The corresponding Pauli
matrices~$S_\alpha^\beta$ are labelled with the vertex index $\alpha$ and the chain number 1, 2, or
3. An example is depicted in \figref{struc}. 
The spin operators $S_\alpha^{1,2,3}$ at each vertex $\alpha$ are spin components of the ancillary spin at this vertex.

To describe a fermion-spin transformation, we use separate Jordan--Wigner
transformations for each chain~$\alpha$,
\begin{subequations}\label{jwt}
\begin{align}
c_\alpha(j) &= \eta_\alpha \brprodz{\alpha}{j-1} \sigma_\alpha^{-}(j) \\
c_\alpha^\dag(j) &= \eta_\alpha \brprodz{\alpha}{j-1} \sigma_\alpha^{+}(j)~,
\end{align}
\end{subequations}
where $\sigma_\alpha^\pm(j) = \tfrac{1}{2} \left[ \sigma_\alpha^x(j) \pm \I \, \sigma_\alpha^y(j)
\right]$. 
The Klein factors~$\eta_\alpha$, with $\eta_\alpha^2 = 1$, are to be chosen to ensure
proper (anti-)\+commutation relations between operators in different chains; they are discussed
later.
Similar to the standard Jordan-Wigner transformation, these relations ensure that a local quadratic
fermionic Hamiltonian is also a local operator in the spin language.
In particular, a useful corollary of these definitions,
\begin{align}
\sigma_\alpha^z(j) &= 2 c_\alpha^\dag(j) c_\alpha(j) - 1 = 1 - 2 c_\alpha(j) c_\alpha^\dag(j)\,,
\label{sigmaz}
\end{align}
shows  that a (magnetic) field in $z$-direction corresponds to a local chemical
potential at a fermionic site. 

\section{Free fermions and 3-spin couplings}\label{3spin}

To complete the description of the transformation, we need to define the operators $\eta_\alpha$.
For the chains directly at the root, $\beta=1, 2, 3$, we define the transformation exactly
like in Ref.~\cite{qsk}: 
\begin{subequations}\label{eta}
\begin{align}
\eta_\beta &= S_0^\beta\,.
\intertext{For any other chain, denoted by $\alpha \beta$ with the parent chain~$\alpha$ and
$\beta=1, 2$, the following definition applies:}
\eta_{\alpha \beta} &= \eta_\alpha \brprodz{\alpha}{L_\alpha} S_\alpha^\beta~.
\end{align}
\end{subequations}
These definitions satisfy the conditions stated in the previous section. 

Let us now consider various nearest-neighbor quadratic fermionic couplings and their spin
counterparts under the constructed transformation.
Within any one-dimensional chain, the Jordan--Wigner transformation is known to convert local
quadratic fermionic Hamiltonians into local quadratic spin Hamiltonians; the factors $\eta_\alpha^2
= 1$ in Eq.~\eqref{jwt} do not affect this. 
Therefore we will examine only the couplings at the vertices between different chains. 
There are two kinds of vertex couplings: those between a parent and a descendant chain and those between two descendant chains of the same parent. 
A coupling term of the first kind between chains $\alpha$ and $\alpha\beta$ (with $\beta=1, 2$)
has the general form
\begin{subequations}\label{Hab}
\begin{align}
H_{\alpha,\alpha\beta} &= u \, c_\alpha(L_\alpha) \, c_{\alpha\beta}(1) + t \,
c_\alpha^\dag(L_\alpha) \, c_{\alpha\beta}(1) + \hc~,
\intertext{which is transformed, using the relation~\eqref{jwt}, into}
H_{\alpha,\alpha\beta}^\text{S} &= S_\alpha^\beta \, \left[ u \, \sigma_\alpha^{-}(L_\alpha) \,
\sigma_{\alpha\beta}^{-}(1)
	- t \, \sigma_\alpha^{+}(L_\alpha) \, \sigma_{\alpha\beta}^{-}(1) + \hc \, \right]~.
\end{align}
\end{subequations}
A coupling of the second kind between chains $\alpha\beta$ and $\alpha\gamma$
(here $\beta\ne\gamma$; $\beta$ and $\gamma$ can be 1 or 2; at the root, $\alpha$ is empty and
$\beta,\gamma=1,2,3$ with ancillary spin operators $S_0^\beta$) has the general form
\begin{subequations}\label{Habg}
\begin{align}
H_{\alpha\beta,\alpha\gamma} &= u \, c_{\alpha\beta}(1) \, c_{\alpha\gamma}(1)
	+ t \, c_{\alpha\beta}^\dag(1) \, c_{\alpha\gamma}(1) + \hc~,
\intertext{which is similarly mapped to}
H_{\alpha\beta,\alpha\gamma}^\text{S} &= S_\alpha^\beta \, S_\alpha^\gamma \, \left[
	u \, \sigma_{\alpha\beta}^{-}(1) \, \sigma_{\alpha\gamma}^{-}(1)
	+ t \, \sigma_{\alpha\beta}^{+}(1) \, \sigma_{\alpha\gamma}^{-}(1) \right] + \hc \nn
&= S_\alpha^\nu \, \epsilon_{\beta\gamma\nu} \, \left[
	\I \,u \, \sigma_{\alpha\beta}^{-}(1) \, \sigma_{\alpha\gamma}^{-}(1)
	+ \I \, t \, \sigma_{\alpha\beta}^{+}(1) \, \sigma_{\alpha\gamma}^{-}(1) + \hc \, \right]~.
\end{align}
\end{subequations}

Let us note that the transformation described can be generalized to arbitrary tree structures,
beyond binary trees. Indeed, any higher-order vertex (with more that three edges) can
be thought of as built out of three-edge vertices. For instance, \figref{struc} can
be viewed as
a five-edge vertex, which allows us to define the Klein factors for all chains outside of
this figure: 
In that case, the internal chains in \figref{struc} are of length zero and do not contribute
products to the Klein factors, but coupling terms involving more than three spins may appear.

\section{XY spin system and fermionic Kondo model}

In this section, we consider a tree structure of spins with local XY couplings and use the JW
transformation backwards in order to find the corresponding fermionic problems.
For a single 1D chain, the Jordan--Wigner transformation maps these to free fermions.
In order to find the corresponding fermionic Hamiltonian for a tree structure, we use the
generalized Jordan--Wigner transformation defined in Eqs.~\eqref{jwt} and \eqref{eta}.
These involve ancillary spin operators $S_\alpha^\beta$, which commute with local spins
$\sigma(j)$, but not with the fermions $c(j)$. We show below that the original XY spin model is
equivalent to a Kondo-type model on the same tree with one impurity spin per vertex.

To simplify the resulting fermionic Hamiltonians, we
introduce, instead of $S$, other spin operators at
the inner vertices, $\tilde{S}_\alpha^\beta$. 
We define
\begin{align}
S_0^\beta = \tilde{S}_0^\beta
\prod_\sst{\text{chain labels $\gamma$} \\
	\text{not beginning} \\ \text{with $\beta$}}
P_\gamma
\qquad\qquad \text{and} \qquad\qquad
S_\alpha^\beta = \tilde{S}_\alpha^\beta
\prod_\sst{\text{chain labels $\alpha\gamma$} \\
	\text{not beginning} \\ \text{with $\alpha\beta$}}
P_{\alpha\gamma}
	\,,
\end{align}
where we introduced the notation
\begin{align}
P_\alpha &= \prodz{\alpha}{L_\alpha}
\end{align}
for the fermionic parity of chain $\alpha$.
As the products consist of Pauli matrices~$\sigma^z$ only, operators~$S_\alpha^\beta$ inherit the
commutation relations of $\tilde{S}_\alpha^\beta$. 
In other words, $\tilde S$ are spin-1/2 operators, and one can verify that they commute with the
fermionic operators.

Let us illustrate this with the example of \figref{struc}:
\begin{subequations}\label{s0}
\begin{align}
S_0^1 &= \tilde{S}_0^1 \, P_2 P_3 \label{s0x}\\
S_0^2 &= \tilde{S}_0^2 \, \left[ P_1 P_{11} \left( P_{12} P_{121} P_{122} \right) \right]
P_3 \\
S_0^3 &= \tilde{S}_0^3 \, \left[ P_1 P_{11} \left( P_{12} P_{121} P_{122} \right) \right]
P_2
\end{align}
\end{subequations}
(the grouping highlights the tree structure),
\begin{subequations}\label{s1}
\begin{align}
S_1^1 &= \tilde{S}_1^1 \, \left( P_{12} P_{121} P_{122} \right) \\
S_1^2 &= \tilde{S}_1^2 \, P_{11} \label{s1y}\\
S_1^3 &= \tilde{S}_1^3 \, P_{11} \left( P_{12} P_{121} P_{122} \right)
\end{align}
\end{subequations}
and
\begin{subequations}\label{s2}
\begin{align}
S_1^1 &= \tilde{S}_{12}^1 \, P_{122} \\
S_1^2 &= \tilde{S}_{12}^2 \, P_{121} \\
S_1^3 &= \tilde{S}_{12}^3 \, P_{121} P_{122}~.
\end{align}
\end{subequations}
The string (parity) operators guarantee that
$S_\alpha^\beta$ commute with the fermionic operators of all chains. 

Again, the Jordan--Wigner transformation is known to map XY-coupled spins in a 1D chain to free fermions, so we only have to examine the two kinds of vertex couplings, as we did in the preceding section. 
They result in Kondo-like couplings of the fermionic chains:
\begin{subequations}\label{Hab2}
\begin{align}
H_{\alpha,\alpha\beta} &= u \, \sigma_\alpha^{-}(L_\alpha) \, \sigma_{\alpha\beta}^{-}(1)
	+ t \, \sigma_\alpha^{+}(L_\alpha) \, \sigma_{\alpha\beta}^{-}(1) + \hc \\
\longrightarrow H_{\alpha,\alpha\beta}^\text{F} &= S_\alpha^\beta \, \left[ u \, c_\alpha(L_\alpha)
\, c_{\alpha\beta}(1)
	- t \, c_\alpha^\dag(L_\alpha) \, c_{\alpha\beta}(1) + \hc \, \right]
\end{align}
\end{subequations}
and
\begin{subequations}\label{Habg2}
\begin{align}
H_{\alpha\beta,\alpha\gamma} &= u \, \sigma_{\alpha\beta}^{-}(1) \, \sigma_{\alpha\gamma}^{-}(1)
	+ t \, \sigma_{\alpha\beta}^{+}(1) \, \sigma_{\alpha\gamma}^{-}(1) + \hc \\
\longrightarrow H_{\alpha\beta,\alpha\gamma}^\text{F} &= S_\alpha^\beta \, S_\alpha^\gamma \, \left[
	u \, c_{\alpha\beta}(1) \, c_{\alpha\gamma}(1)
	+ t \, c_{\alpha\beta}^\dag(1) \, c_{\alpha\gamma}(1) \right] + \hc \nn
&= S_\alpha^\nu \, \epsilon_{\beta\gamma\nu} \, \left[
	\I \,u \, c_{\alpha\beta}(1) \, c_{\alpha\gamma}(1)
	+ \I \, t \, c_{\alpha\beta}^\dag(1) \, c_{\alpha\gamma}(1) + \hc \, \right]~. 
\end{align}
\end{subequations} 
Thus, inter-chain couplings are controlled by the ancillary spins.


\section{Majorana braiding and the spin representation}
\label{sec:MajBraid}

In this section, we are interested in spin implementation of free-fermion models on tree structures.
These can be applied, in particular, to realize (physically simulate) Majorana qubits and quantum logical operations
using ordinary quantum bits.

Majorana modes arising in the topological phase of the Kitaev chain~\cite{kit}, a one-dimensional
fermionic system, can be braided in a $T$-junction geometry by local tuning of the chemical
potential \cite{1dw}. 
One can see from the discussion above that similar to Refs.~\cite{qsk,emb}, the corresponding
spin model involves Ising couplings within the chains, the ancillary-spin-controlled Ising
couplings at the junctions as well as a transverse magnetic field.

In the following, the spin indices are swapped for convenience, to ensure the resulting $zz$ Ising
couplings and the transverse field in the $x$ direction. 
Furthermore, we use fermionic Majorana operators $\gamma_\alpha(m)$~\footnote{%
The Majorana operators are connected to the usual fermionic creation and annihilation operators
$c^\dag, c$ in the following way: $\gamma_\alpha(2j-1) = c_\alpha(j) + c_\alpha^\dag(j)$ and
$\gamma_\alpha(2j) = -\I \left[ c_\alpha(j) - c_\alpha^\dag(j) \right]$.}
satisfying the anti-commutation relations,
\begin{align}
\{ \gamma_\alpha(m), \gamma_\beta(n) \}_+ &= 2 \, \delta_{\alpha\beta} \, \delta_{mn}\,,
\end{align}
to express the transformation in a convenient form:
\begin{subequations}\label{jwtM}
\begin{align}
\gamma_\alpha(2j-1) &= \eta_\alpha \prodx{\alpha}{j-1} \sigma_\alpha^z(j) \\
\gamma_\alpha(2j) &= \eta_\alpha \prodx{\alpha}{j-1} \sigma_\alpha^y(j) \\[0.4cm]
\Rightarrow \sigma_\alpha^x(j) &= \I \, \gamma_\alpha(2j-1) \, \gamma_\alpha(2j)~.
\end{align}
\end{subequations}
The Klein factors $\eta_\alpha$ are those defined in Equations~\eqref{eta}. 

The transformation relates the topological (nontopological) phase in the fermionic chains to the ferromagnetic (paramagnetic) phase of the spin system (for more details see Appendix). 
Now we can simply translate into the spin system the unitary operator produced by, \eg, counter-clockwise braiding of Majorana modes $\gamma_\text{A}, \gamma_\text{B}$ \cite{1dw}:
\begin{align}
U &= \exp\lft( \tfrac{\pi}{4} \, \gamma_\text{A}  \, \gamma_\text{B} \right)
\end{align}
(how this can be implemented may depend on the tree structure and the initial positions of
$\gamma_A$ and $\gamma_B$).
In the case of two Majorana modes that are provided by one topological segment located in a single
chain before and after the braiding, the Klein factors cancel in the spin representation, so the
additional spin mediating the coupling at the junction does not influence the result of the
operation and is left unaffected at the end~\cite{emb}. 

Braiding neighbouring Majorana modes from two topological segments in different chains corresponds
to a more complicated operation in the spin system. 
By choosing, \eg, $\gamma_\text{A} = \gamma_1(2m-1)$ and $\gamma_\text{B} = \gamma_3(2n-1)$, we obtain:
\begin{align}
U_{1,3} &= \exp\lft[ \tfrac{\pi}{4} \cdot S_0^x \prod_{j=1}^{m-1} \sigma_1^x(j) \cdot \sigma_1^z(m)
	\cdot S_0^z \prod_{k=1}^{n-1} \sigma_3^x(k) \cdot \sigma_3^z(n) \right] \nn
&= \exp\lft[ -\I \, \tfrac{\pi}{4} \, S_0^y \cdot \sigma_1^z(m) \prod_{j=1}^{m-1} \sigma_1^x(j)
	\prod_{k=1}^{n-1} \sigma_3^x(k) \cdot \sigma_3^z(n) \right] \nn
&\xrightarrow{\text{effectively}} \exp\lft[ -\I \, \tfrac{\pi}{4} \, S_0^y \, \sigma_1^z(m) \, \sigma_3^z(n) \right]~, \label{majRot}
\end{align}
if the spins outside the ferromagnetic intervals are polarised in the $x$-direction.
When expressed in terms of the Pauli matrices $\tau$ for the two 'topological' qubits involved (two
ferromagnetic intervals, cf.~Ref.~\cite{emb}), this gives
\begin{equation}\label{U13}
U(1,3) = \exp\lft[ -\I \, \tfrac{\pi}{4} \, S_0^y \, \tau_1^z \, \tau_2^z \right] \,.
\end{equation}
A detailed description of this operation in the spin language is given in the Appendix.

However, for two `topological' intervals in the same chain
we obtain a similar expression, but without the intermediate ancillary spin:
\begin{equation}
U = \exp\lft[ -\I \, \tfrac{\pi}{4} \tau_1^y \, \tau_2^z \right] \,.
\end{equation}
Thus one obtains a two-qubit operation.

In a more general situation with arbitrary initial position of two distant braided boundaries (and
associated Majorana modes), the brading operation involves, apart from these two qubits, the
ancillary spins at all intermediate vertices as well as the parity (qubit-flip) operators $\prod
\sigma_x$ for all intermediate qubit intervals. Thus, the braiding implements not a two-qubit
operation but a multi-qubit operation (and also entangles qubits with the ancillas).

Here a few comments are in order: First, to achieve direct two-qubit gates between distant
`topological' qubit intervals, one can complement the described braiding operation with further
operations involving intermediate qubits. However, for the purposes of quantum
computation one does not necessarily need two-qubit logic gates between distant qubits since
two-qubit gates between neighbors are sufficient, as they form a universal set of gates together
with single-qubit operations.
Furthermore, one can also view this subtlety from a different perspective.
Instead of thinking in terms of the qubit description, one can describe the operations in terms of
the fermionic (Majorana) modes involved. Then the braiding operations implement two-fermion gates,
and one deals with {\it fermionic quantum computation}. This viewpoint may be useful for simulations
of fermionic Hamiltonians (see e.g. Ref~\cite{AspuruGuzik}), including many-body solid-state models 
and complex individual molecules.

A further remark concerns the symmetry and the braiding procedure:
Each of the chains considered belongs to the BDI symmetry class~\cite{AltlandZirnbauer},
with a time-reversal-type symmetry ${\cal T}$ such that ${\cal T}^2=1$. In a single
chain~\eqref{eq:chain}, a Majorana zero mode appears at each boundary between
topological and non-topological regions.
A vertex connecting three chains can be viewed as an edge of a 1-D system.
Here the symmetry becomes crucial~\cite{FidKitaev10,Meidan}.
If the ${\cal T}$ symmetry is preserved by the chain coupling at the vertex, the
edge (vertex) carries an integer ($\mathbb{Z}$) `topological' charge. In our case this allows for
configurations with more than one Majorana zero mode at the vertex and an unwanted extra degeneracy
when during the braiding procedure this vertex connects two or three `topological' regions.
A $\cal T$-breaking chain coupling, however, places the system to the D class
with a $\mathbb{Z}_2$ invariant, and typically one (or no) Majorana zero mode exists at the
vertex (cf.~Ref.~\cite{1dw}). In this case, no extra degeneracies arise during the braiding
operations. In particular, this is the case for the coupling considered in Ref.~\cite{emb}.

\section{Discussion}

The Jordan--Wigner transformation maps free fermionic Hamiltonians to local spin Hamiltonians.
Furthermore, a nearest-neighbor hopping term is mapped to a local quadratic spin term.
Some generalizations of the Jordan--Wigner transformation to higher-dimensional lattices were
proposed~\cite{Fradkin,YWang,HuertaZanelli,SWang}, which, however, map a local hopping fermionic
term to a spin term involving many, often infinitely many, spins on distant sites.
Other approaches (\eg, Ref.~\cite{VerstraeteCirac}) use ancillary degrees of freedom but also map
free fermionic terms to fourth- or higher-order spin terms.

On one hand, this implies that only some unusual spin models may be analyzed with the help of such
transformations. On the other hand, one could use a spin--fermion mapping to implement a fermionic
model in a system built of spins (or qubits). With the motivation to implement fermionic (Majorana)
degrees of freedom in a realistic qubit system, we have extended an earlier result of Ref.~\cite{qsk} to
arbitrary tree structures. The resulting transformation maps nearest-neighbor fermionic terms to
nearest-neighbor spin terms. Thus, it allows for an implementation of the Majorana physics in tree
structures built out of qubit chains, extending the results of Ref.~\cite{emb}.
This transformation provides, \eg, a spin equivalent of Majorana braiding operations.
We have further shown that this construction can be generalized to arbitrary tree structures.

It must be noted that these mappings involve an enlargement of the original Hilbert space, due to the addition of spins $\tilde{S}_\alpha$ to the system. Thus, the degeneracy of all states is multiplied by a factor of 2 to the power of the number of inner vertices, but the accuracy of the mapping is not affected. 

Finally, we would like to mention that experimental realizations of the 3-spin
interactions crucial for our mapping were discussed in the literature (see, \eg,
Refs.~\cite{PhysRevX.3.041018},\cite{emb}).

\section{Acknowledgements}

We are grateful to Guido Pagano for useful discussions. 
This research was financially supported by the DFG-RSF grant No. 16-42-01035 (Russian node) and No. SH 81/4-1, MI 658/9-1  (German node).

\appendix*
\section{Spin representation of two-interval Majorana braiding}\label{app}

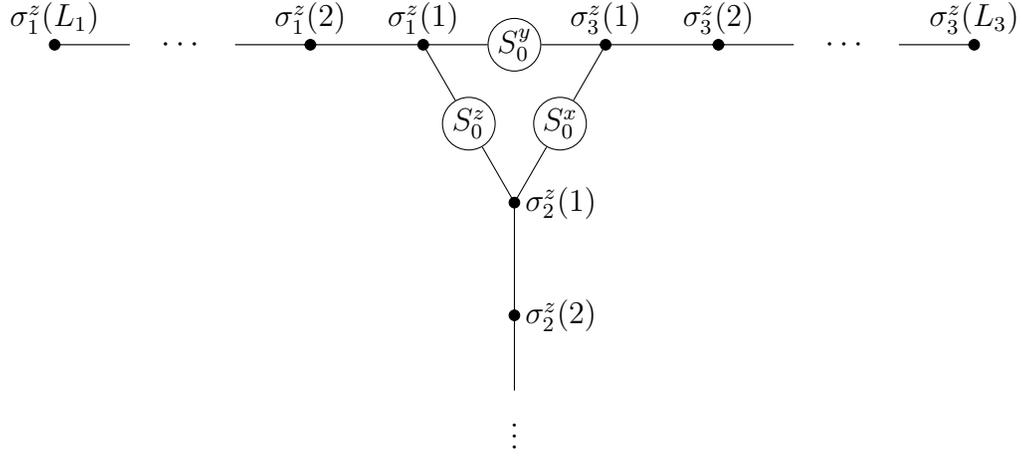
\begin{figure}[hbt]
\large
\newcommand*{\kreis}{2pt}
\newcommand*{\lang}{1.5}
\newcommand*{\mitt}{1}
\newcommand*{\rad}{0.7}
\pgfmathsetmacro{\radd}{2*\rad}
	\begin{tikzpicture}[kop/.style={circle,draw,inner sep=0pt,minimum size=7mm}]
	\path (0, \rad) node[kop] (kopZ) {} node {$S_0^y$}
		arc(90:210:\rad) node[kop] (kopY) {} node {$S_0^z$}
		arc(210:330:\rad) node[kop] (kopX) {} node {$S_0^x$};
	\path (0, -\radd) coordinate (s3)
		arc(270:30:\radd) coordinate (s1)
		arc(30:150:\radd) coordinate (s2);
	\filldraw (kopZ) -- (s2)
		circle (\kreis) node[above] {$\sigma_1^z(1)$} -- ++(-\lang, 0)
		circle (\kreis) node[above] {$\sigma_1^z(2)$} -- ++(-\mitt, 0)
		++(-\rad, 0) node {$\cdots$} ++(-\rad, 0) -- ++(-\mitt, 0)
		circle (\kreis) node[above] {$\sigma_1^z(L_1)$};
	\filldraw (kopZ) -- (s1)
		circle (\kreis) node[above] {$\sigma_3^z(1)$} -- ++(\lang, 0)
		circle (\kreis) node[above] {$\sigma_3^z(2)$} -- ++(\mitt, 0)
		++(\rad, 0) node {$\cdots$} ++(\rad, 0) -- ++(\mitt, 0)
		circle (\kreis) node[above] {$\sigma_3^z(L_3)$};
	\filldraw (s3)
		circle (\kreis) node[right] {$\sigma_2^z(1)$} -- ++(0, -\lang)
		circle (\kreis) node[right] {$\sigma_2^z(2)$} -- +(0, -\mitt)	
		++(0, -\lang) node {$\vdots$};
	\draw (s2) -- (kopY) -- (s3) -- (kopX) -- (s1);
	\end{tikzpicture}
\caption{Ising spin chains in a $T$-geometry. A fermionic $T$-junction suitable for Majorana
braiding~\cite{1dw} has a spin representation of this structure~\cite{emb}, which is described by
the Hamiltonian in Eq.~\eqref{spinH}. 
The couplings between three Ising spin chains are mediated by the components of an additional spin~$S_0$, cf. Ref.~\cite{qsk} and Section~\ref{3spin}. 
The system can be manipulated by tuning transverse fields (not depicted here) that act on the individual spins~$\sigma_\alpha(j)$ of the three chains. 
}
\label{spinT}
\end{figure}

To obtain a spin representation for a fermionic $T$-junction of Kitaev chains, we use the
transformation given in Eqs.~\eqref{jwtM}. 
This yields \cite{emb}:
\begin{subequations}\label{spinH}
\begin{align}
H &= \sum_{\alpha=1}^3 H_{0, \alpha} + H^S_\text{int} \\
H_{0, \alpha} &= -\sum_{j=1}^{L_\alpha} h_\alpha(j) \, \sigma_\alpha^x(j)
	- J \sum_{j=1}^{L_\alpha-1} \sigma_\alpha^z(j) \, \sigma_\alpha^z(j+1) \label{eq:chain}\\
H^S_\text{int} &= -\frac{1}{2} \sum_{\alpha\beta\gamma}^{\phantom{L_\alpha}} |\epsilon^{\alpha\beta\gamma}|
	\, J_{\alpha\beta} \, S_0^\gamma \, \sigma_\alpha^z(1) \, \sigma_\beta^z(1)~, \label{spinHint}
\end{align}
\end{subequations}
a system of Ising spin chains with a local transverse magnetic field~$h_\alpha(j)$, which corresponds to the locally tunable chemical potential in the fermionic system. 
Assuming $J > 0$, any interval of spins with $h \gg J$ in one of the chains is ferromagnetic, whereas $h \ll J$ results in a trivial (paramagnetic) phase. 
The three chains are linked by the components of an additional central spin~$S_0$ via 3-spin couplings of strength~$J_{\alpha\beta} = J_{\beta\alpha}$. 
This structure is depicted in \figref{spinT}. 

Now we consider the spin equivalent of braiding Majorana modes from two different topological
intervals in the fermionic system. 
The topological intervals and adiabatic shifts of their boundaries within a chain can be translated
to the ferromagnetic intervals in the spin representation exactly as for a single
topological interval~\cite{emb}: 
The fermionic-parity groundstates $\left| 0 \right>, \left| 1 \right>$ of a topological interval correspond to linear combinations
\begin{subequations}\label{parStates}
\begin{align}
\ket{0} &\equiv \frac{\ket{\up\up\up} + \ket{\dn\dn\dn}}{\sqrt{2}}~,
\\
\ket{1} &\equiv \frac{\ket{\up\up\up} - \ket{\dn\dn\dn}}{\sqrt{2}}
\end{align}
\end{subequations}
of the ferromagnetic spin eigenstates. 
However, two-interval braiding cannot be effected in such a way that at most one of the 3-spin couplings in Equation~\eqref{spinHint} is relevant at each step. 
Therefore, the \emph{coupler} spin~$S_0$ undergoes a generally non-trivial rotation in the process,
which we will examine in the following. 

Initially, the topological/ferromagnetic intervals have to be prepared, \eg, in the first and
third chain at some distance from the coupler spin~$S_0$. 
We assume that $J_{12} = J_{13} = J_{23} > 0$ for illustration. 
First, we consider an initial state with the spins in both intervals and the coupler aligned
in the $+z$-direction:
\begin{align}
\ket{\psi_0} &= \ket{\up\up\up \!\odot}_1
	\otimes \ket{\odot \odot}_2
	\otimes \ket{\odot\! \up \up \up}_3
	\otimes \ket{S_0 \uparrow}\,.
\intertext{Here the indices denote the three spin chains; the corresponding arrows indicate the spin orientation in ferromagnetic ($\up$/$\dn$) and paramagnetic areas ($\odot$). 
They symbolize the locations of the ferromagnetic intervals in the $T$-junction geometry~(\figref{spinT}), but
the calculation does not depend on the specific interval lengths and distances to the
coupler.
The first step comprises shifting the right boundary of the first interval (\ie, one Majorana mode
in the fermionic system) into the second chain, which results in the state}
\ket{\psi_1} &= \ket{\odot\! \up \up \up}_1
	\otimes \ket{\up \up}_2
	\otimes \ket{\odot\! \up \up \up}_3
	\otimes \ket{S_0 \uparrow}\,.
	\label{eq:psi1}
\end{align}
With $\ket{S_0 \downarrow}$ as initial coupler state, the spins in the second chain in
Eq.~(\ref{eq:psi1}) would just be flipped compared to those in the first chain. 

The non-trivial part begins when the second ferromagnetic interval is also shifted to the junction, 
while the spin orientation of the ferromagnetic intervals remains fixed.
One can verify that for any initial state, including superpositions, the final state at this stage
is always the same as in the case when the second ferromagnetic interval is adiabatically
shifted towards the junction at $J_{13}=J_{23}=0$, and only then these couplings are
slowly turned on. This observation simplifies the further calculation.
Indeed, at the end of this stage, the coupler spin rotates
to adjust to the change of its effective magnetic field from the $z$-direction to the space diagonal
$\frac{1}{\sqrt{3}} \, \vektor{1}{1}{1}$:
\begin{align}
\left| \psi_2 \right> &=
\left| \odot\! \up \up \up \right>_1
	\otimes \left| \up \up \right>_2
	\otimes \left| \up \up \up \!\odot \right>_3
	\otimes \left[ \cos\frac{\varphi}{2} \left| S_0 \uparrow \right>
\label{psi2}
	+ \e^{\I \tfrac{\pi}{4}} \sin\frac{\varphi}{2} \left| S_0 \downarrow \right>
\right] \,,
\end{align}
where $0<\varphi<\pi/2$ and $\cos\varphi=1/\sqrt{3}$.
Similarly, at the next stage, when the ferromagnetic interval in the first chain is shifted away from
the junction, the coupler spin adjusts to the $x$-direction:
\begin{align}\label{eq:psi3}
\left| \psi_3 \right> &= 
\left| \up \up \up \!\odot
\right>_1
	\otimes \left| \up \up \right>_2
	\otimes \left| \up \up \up \!\odot \right>_3
	\otimes \tfrac{1}{\sqrt{2}} \big[ \left| S_0 \uparrow \right> + \left| S_0 \downarrow \right> \big]~.
\end{align}
Retracting the remaining ferromagnetic interval back to the third chain is a trivial step again:
\begin{align}\label{eq:psi4}
\left| \psi_4 \right> &=  
\left| \up \up \up \!\odot
\right>_1
	\otimes \left| \odot \odot \right>_2
	\otimes \left| \odot\! \up \up \up \right>_3
	\otimes \tfrac{1}{\sqrt{2}} \big[ \left| S_0 \uparrow \right> + \left| S_0 \downarrow
\right> \big]~.	
\end{align}
Unlike in the case of single-interval braiding~\cite{emb}, the coupler spin does not return to its
intial state at the end of the operation (see, however, discussion in Sec.~\ref{sec:MajBraid}). 
In Eqs.~(\ref{eq:psi3}), (\ref{eq:psi4}), we have dropped the overall phase factor that can be linked to the geometric phase of
the spin evolution. It turns out to be the same for all states of interest to us
(cf.~Eq.\eqref{braidResult} below) and will be omitted.

For other initial conditions, the operation can be treated similarly, giving the complete result
\begin{subequations}\label{braidResult}
\newcommand*{\chisp}{\; \blacklozenge \blacklozenge \blacklozenge \;}
\newcommand*{\psisp}{\; \lozenge \lozenge \lozenge \;}
\begin{align}
\left| \chisp \!\odot \right>_1
	\otimes \left| \odot\! \chisp \right>_3
	\otimes \left| S_0 \uparrow \right>
&\quad\longrightarrow\quad \left| \chisp \!\odot \right>_1
	\otimes \left| \odot\! \chisp \right>_3
	\otimes \tfrac{1}{\sqrt{2}} \big[ \left| S_0 \uparrow \right> + \left| S_0 \downarrow \right> \big] \\
\left| \chisp \!\odot \right>_1
	\otimes \left| \odot\! \psisp \right>_3
	\otimes \left| S_0 \uparrow \right>
&\quad\longrightarrow\quad \left| \chisp \!\odot \right>_1
	\otimes \left| \odot\! \psisp \right>_3
	\otimes \tfrac{1}{\sqrt{2}} \big[ \left| S_0 \uparrow \right> - \left| S_0 \downarrow \right> \big] \\
\left| \chisp \!\odot \right>_1
	\otimes \left| \odot\! \chisp \right>_3
	\otimes \left| S_0 \downarrow \right>
&\quad\longrightarrow\quad - \left| \chisp \!\odot \right>_1
	\otimes \left| \odot\! \chisp \right>_3
	\otimes \tfrac{1}{\sqrt{2}} \big[ \left| S_0 \uparrow \right> - \left| S_0 \downarrow \right> \big] \\
\left| \chisp \!\odot \right>_1
	\otimes \left| \odot\! \psisp \right>_3
	\otimes \left| S_0 \downarrow \right>
&\quad\longrightarrow\quad \left| \chisp \!\odot \right>_1
	\otimes \left| \odot\! \psisp \right>_3
	\otimes \tfrac{1}{\sqrt{2}} \big[ \left| S_0 \uparrow \right> + \left| S_0 \downarrow \right> \big]
\end{align}
\end{subequations}
with placeholders $\{ \blacklozenge, \lozenge \} = \{ \up, \dn \}$.
The initial as well as final state of the second chain is always $\left| \odot \odot \right>_2$. 
Using the parity eigenstates~\eqref{parStates}, we can verify that Eqs.~\eqref{braidResult} indeed
correspond to the Majorana braiding. For instance,
\begin{align}
&\left| 0 \right>_1 \otimes \left| 0 \right>_3 \otimes \left| S_0 \uparrow \right> = \frac{\uket_1 + \dket_1}{\sqrt{2}}
	\otimes \frac{\uket_3 + \dket_3}{\sqrt{2}} \otimes \left| S_0 \uparrow \right> \nn
&\longrightarrow \frac{\uket_1 \otimes \uket_3 + \dket_1 \otimes \dket_3}{2}
	\otimes \frac{\left| S_0 \uparrow \right> + \left| S_0 \downarrow \right>}{\sqrt{2}} \nn
&\qquad\: {} + \frac{\uket_1 \otimes \dket_3 + \dket_1 \otimes \uket_3}{2}
	\otimes \frac{\left| S_0 \uparrow \right> - \left| S_0 \downarrow \right>}{\sqrt{2}} \nn
&= \tfrac{1}{\sqrt{2}} \big[ \left| 0 \right>_1 \otimes \left| 0 \right>_3 \otimes \left| S_0 \uparrow \right>
	+ \left| 1 \right>_1 \otimes \left| 1 \right>_3 \otimes \left| S_0 \downarrow \right> \big]~.
\end{align}
In addition to the ferromagnetic intervals, the superposition involves the coupler spin, in accordance with the expressions~\eqref{majRot}, \eqref{U13}, which are compatible with the results~\eqref{braidResult}. 

Note that the coupling $J_{13}$ is not necessary for the operation we considered here. 
The choice of $J_{13} = 0$ simplifies the coupler rotations and leads to the same results.

\bibliography{trees3}

\end{document}